\documentstyle[12pt]{article}
\begin{document}
\begin{center}
{\large {\bf \mbox{EXACT SOLUTIONS OF THE DIRAC EQUATION}\\[0pt]
\mbox{FOR MODIFIED COULOMBIC POTENTIALS}}}\\[0pt]
\vspace{0.5in} ANTONIO SOARES DE CASTRO$^{a}$ AND JERROLD FRANKLIN$^{b}$\\[%
0pt]
\vspace{0.1in} $^{a}${\it DFQ, UNESP, C.P. 205, 12500-000 Guaratinguet\'{a}
SP, Brasil}\\[0pt]
$^{b}${\it Department of Physics, Temple University, Philadelphia, PA
19122-6082}\\[0pt]
\vspace{.3in}
Received\hspace{.3in} February 2000
\end{center}
\vspace{.1in}
\begin{quote}
{\small Exact solutions are found to the Dirac equation for a combination of
Lorentz scalar and vector Coulombic potentials with additional non-Coulombic
parts. An appropriate linear combination of Lorentz scalar and vector
non-Coulombic potentials, with the scalar part dominating, can be chosen to give
exact analytic Dirac wave functions.  The method works for the ground state or
for the lowest orbital state with \mbox{$l=j-\frac{1}{2}$,} for any $j$.}
\end{quote}
\vspace{.3in}

     In a previous letter\cite{jf}, simple exact solutions were found for the
Dirac equation for the combination of a Lorentz vector Coulomb potential with a
linear confining potential that was a particular combination of Lorentz scalar
and vector parts.  In this letter, we extend the method of Ref.\cite{jf} to the
more general case of an arbitrary combination of Lorentz scalar and vector
Coulombic potentials with a particular combination of Lorentz scalar and vector
non-Coulombic potentials, $S(r)$ and $V(r)$.  The non-Coulombic potentials can
have arbitrary radial dependence, but must be related by
\begin{equation}
V(r)=-\frac{E}{m}S(r),
\end{equation}
where $E$ is the bound state energy for a particle of mass $m$.
This is the same relation between the vector and scalar non-Coulombic potentials
as was required in Ref.\cite{jf}

The Dirac equation we solve is
\begin{equation}
\left[{\bf\alpha\cdot p}+\beta m -\frac{(\lambda+\beta\eta)}{r} +V(r)+\beta
S(r)\right]\psi=E\psi,
\end{equation}
where {\boldmath$\alpha$\unboldmath} and $\beta$ are the usual Dirac matrices.
The four component wave function $\psi$ can be written in terms of two component
spinors $u$ and $v$ as
\begin{equation}
\psi =  N\left(\begin{array}{c}u\\ v\end{array}\right),
\end{equation}
with $N$ an appropriate normalization constant.
The two component spinors $u$ and $v$ satisfy the equations
\begin{eqnarray}
({\bf\sigma\cdot p})v & = &
\left[E-m+\frac{(\lambda+\eta)}{r}-V(r)-S(r)\right]u\\
({\bf\sigma\cdot p})u & = &
\left[E+m+\frac{(\lambda-\eta)}{r}-V(r)+S(r)\right]v.
\end{eqnarray}

The key step in generating relatively simple exact solutions of the Dirac
equation is to choose a particularly simple form for the function $v(r)$
\begin{equation}
v(r)=i\gamma({\bf\sigma\cdot{\hat r}})u(r),
\label{eq:v}
\end{equation}
where $\gamma$ is a constant factor, to be determined by the solution to the
Dirac equation.  This is the form of $v(r)$ that was found in Ref.\cite{jf} for
a Coulomb plus linear confining potential.  This ansatz for $v(r)$ has also been
used as the basis for generating approximate saddle point solutions for the
Dirac\cite{sp1} and Breit\cite{sp2} equations.

For now, we limit our discussion to spherically symmetric states $u(r)$ that
depend only on the radial coordinate.  We will discuss orbitally excited states
later in this paper. Using the form of $v(r)$ given by Eq.\ (\ref{eq:v}),
equations (4) and
(5) reduce to two first order ordinary diffential equations for $u(r)$
\begin{eqnarray}
\frac{du}{dr} &
= &\frac{1}{\gamma}\left[E-m-V(r)-S(r)
+\frac{(\lambda+\eta-2\gamma)}{r} \right]u(r)\\
\frac{du}{dr} &
= & -\gamma\left[E+m-V(r)+S(r)+\frac{(\lambda-\eta)}{r}\right]u(r).
\end{eqnarray}
Equations (7) and (8) are two independent equations for the same quantity, so
that each term in one equation can be equated with the corresponding term
in the other equation having the same radial dependence.  This leads to
\begin{equation}
\gamma^2=\frac{m-E}{m+E}=\frac{S(r)+V(r)}{S(r)-V(r)}
=\frac{\lambda+\eta-2\gamma}{\eta-\lambda}.
\label{gs}
\end{equation}

The relations in Eq.\ (9) can be rearranged, after some algebra, to give
\begin{equation}
\gamma=\frac{\lambda+\eta}{1+b},
\label{gb}
\end{equation}
with
\begin{equation}
b=\pm\sqrt{1-\lambda^2+\eta^2}.
\label{b}
\end{equation}
The constant b can have either sign.  Although $b$ must be positive in the pure
Coulombic case, we will see that a negative $b$ is possible if
the Lorentz scalar potential $S(r)$ is more singular at the origin than $1/r$.
The bound state energy can be written as
\begin{equation}
E=m\left(\frac{1-\gamma^2}{1+\gamma^2}\right)
=m\left(\frac{b\lambda-\eta}{\lambda-b\eta}\right),
\label{e}
\end{equation}
or as
\begin{equation}
E=-m\frac{V(r)}{S(r)}.
\label{v}
\end{equation}

The wave function $u(r)$ can be found by solving
differential equation (8) to give
\begin{equation}
u(r)=r^{b-1}\exp\left[-a\left(r +\frac{1}{m}\int S(r)dr\right)\right],
\label{u}
\end{equation}
where the constant a is given by
\begin{equation}
a=\gamma(m+E)=\pm\sqrt{m^2-E^2}.
\label{a}
\end{equation}
The constants $\gamma$ and $a$ could be negative if $S(r)$
approaches a large enough negative constant or diverges negatively as $r$
becomes infinite.  The integral in equation (\ref{u}) can diverge at the origin
or as $r\rightarrow\infty$, or for any finite $r$, as long as the quantity in
square brackets in Eq.\ (\ref{u}) remains negative.

The constraint equation
(\ref{v}) shows that in order for this class of exact solutions to apply, the
Lorentz vector and scalar non-Coulombic potentials must have the same radial
dependence.  As long as this constraint is satisfied, the results in
Eqs.\ (10)-(15) represent a complete exact solution for the wave function and
bound state energy of the Dirac Hamiltonian given in equation (2).

We note from Eq.\ (\ref{e}) that the energy seems to depend only on the
Coulombic coupling constants $\lambda$ and $\eta$, and does not seem to depend
on the
non-Coulombic potentials.  However, we could alternatively say from Eq.\
(\ref{v})  that the energy depends only on the ratio of the non-Coulombic
potentials
$V(r)$ and $S(r)$, and does not seem to depend on the Coulombic coupling
constants.
Then, Eq.\ (\ref{e}) could be considered a constraint equation on the Coulombic
coupling constants.  The actual situation is that the energy does depend on all
the potentials but, because of the severe constraints imposed by equations
(\ref{e}) and (\ref{v}) taken together, the energy can be written in terms of
one set of potentials or the other.
Although the possibility of this class of exact solutions is limited by the
constraints on the potentials, this still permits a wide range of non-Coulombic
potentials.

We now consider conditions imposed on the potentials and the wave function
parameters by the physical requirements that the potentials be real and the wave
function normalizable.  We see from Eq.\ (\ref{gs}) that $\gamma$ must be real,
and then from Eq.\ (\ref{gb}) that $b$ must be real.  This requires the
Coulombic potentials to satisfy the condition
\begin{equation}
1-\lambda^2+\eta^2 \ge 0.
\end{equation}
The reality of $\gamma$ restricts possible bound state energies to the range
\begin{equation}
-m < E < m.
\end{equation}
Note that negative energies can occur, but $E+m$ cannot be negative.
Also, $E$ cannot equal $\pm m$, because this would lead to an unormalizable wave
function.  This condition on $E$, along with Eq.\ (\ref{v}), means that
$V(r)$ must always be less in magnitude than $S(r)$.

We discuss the remaining conditions on the parameters in terms of three
sub-classes of solution:
\begin{enumerate}
\item The "normal" class of solutions has $b$, $\gamma$, and $a$ all positive.
In this case, we see from Eq.\ (\ref{gb}) that the Coulombic potentials must
satisfy the further condition
\begin{equation}
\lambda + \eta > 0.
\end{equation}
\item Sub-case 2 has $b$ negative, with $\gamma$ and $a$ still positive.
The constant $b$ can be negative
if the product $aS(r)$ is positively divergent at the origin faster than $1/r$.
Then each of Eqs.\ (10)-(15) holds just as for positive $b$, and the wave
function is still normalizable.  Sub-case 1 with positive $b$ transforms
smoothly into the pure Coulombic solution as the non-Coulombic potential tends
to zero everywhere.  But this is not true for sub-case 2 with $b$ negative.
This sub-case requires the non-Coulombic potential to be dominant at the origin,
and so has no corresponding pure Coulombic limit.
\item  Sub-case 3 has a negative $\gamma$ and a negative $a$, while $b$ can have
either sign,
as discussed in sub-cases 1 and 2 above.
A negative $a$ is possible if the non-Coulombic potential diverges or approaches
a constant as $r\rightarrow\infty$, so that the integral in Eq.\ (\ref{a})
diverges faster than $r$ at large $r$.  Since $a$ is negative, the potential
$S(r)$ must be {\em negative}  at large $r$.
Then all of Eqs.\ (10)-(15) hold as for positive $a$, and the wave function is
still normalizable.  This case is highly unusual, because it allows the
possibility of a potential that is negative everywhere and diverges negatively
at both the origin and infinite $r$.
We know of no other example in quantum mechanics where a potential
that diverges negatively at infinity can lead to a normalizable ground state.
The reason this is possible here can be seen from
Eqs.\ (7) and (8).  There it is seen that $S(r)$  enters the differential
equations for $u(r)$ only in the combinations $\gamma S(r)$ or $S(r)/\gamma$.
Since these effective potentials are positive, the resulting wave function is
normalizable.  As with sub-case 2, the case with $\gamma$ and $a$ negative
cannot
not approach a pure Coulombic case because the non-Coulombic potential must be
dominant at large r.
\end{enumerate}

We now look at some special cases.  If the non-Coulombic potentials are absent,
then the solutions are for a general linear combination of Lorentz vector and
Lorentz scalar Coulombic potentials.  If either constant, $\lambda$ or $\eta$,
is zero, we recover the usual solutions of the Dirac equation for a
pure scalar or vector
Coulombic potential.  The Coulombic potentials cannot both be absent
(while keeping a non-Coulombic part) because then $\gamma$ would be zero
and b one, leading to a constant, unnormalizable wave fucntion.

For a power law non-Coulombic potential of the form
\begin{equation}
S(r)=\mu(s+1) r^s,\quad s\ne -1,
\end{equation}
the wave function will be given by
\begin{equation}
u(r)=r^{b-1}exp\left[-a\left(r -\frac{\mu}{m}r^{s+1}\right)\right].
\label{us}
\end{equation}
All three sub-classes are possible for this wave function, depending on the
ranges of the parameters $b$, $a$, $\mu$, and $s$.

The method does not work for radially excited states, because the simple ansatz
of Eq.\ (\ref{eq:v}) for $v(r)$ does not lead to consistent equations for
$du/dr$ in that case.  But, the method does work for the lowest orbitally
excited state for which $l=j-\frac{1}{2}$.
This can be seen by writing the Dirac wave function in terms of  radial
functions \mbox{$f(r)$ and $g(r)$}
\begin{equation}
\psi =  N\left(\begin{array}{c}
f(r){\cal Y}^j_{j-\frac{1}{2}}\\
-ig(r){\cal Y}^j_{j+\frac{1}{2}}
\end{array}\right),
\end{equation}
where ${\cal Y}^j_l$ is a two component angular spinor function corresponding to
total angular momentum $j$ and orbital momentum $l$.
The radial functions satisfy the equations
\begin{eqnarray}
\left(\frac{d}{dr}+\frac{1-\kappa}{r}\right)f & = &
-\left[E+m+\frac{(\lambda-\eta)}{r}+S(r)-V(r)\right]g\\
\left(\frac{d}{dr}+\frac{1+\kappa}{r}\right)g & = &
\left[E-m+\frac{(\lambda+\eta)}{r}-S(r)-V(r)\right]f,
\end{eqnarray}
where $\kappa=j+\frac{1}{2}$ is the principal quantum number of the state.

For the radial function $g(r)$, we make the ansatz
\begin{equation}
g(r)=\gamma f(r).
\end{equation}
Then, equations (22) and (23) reduce to two first order differential equations
for $f(r)$
\begin{eqnarray}
\frac{df}{dr} &= & -\gamma\left[E+m-V(r)+S(r)
+\frac{(\lambda-\eta)}{r} -\frac{(\kappa-1)}{\gamma r} \right]f(r)\\
\frac{df}{dr} &= &\frac{1}{\gamma}\left[E-m-V(r)-S(r)
+\frac{(\lambda+\eta)}{r} -\frac{\gamma(\kappa+1)}{r} \right]f(r).
\end{eqnarray}

Equating the corresponding terms having the same radial dependence in these two
equations results in
\begin{equation}
\gamma^2=\frac{m-E}{m+E}=\frac{S(r)+V(r)}{S(r)-V(r)}
=\frac{\lambda+\eta-2\gamma\kappa}{\eta-\lambda}.
\label{gl}
\end{equation}
Equation (\ref{gl}) is the same as Eq.\ (\ref{gs}) with the replacements
\begin{equation}
\lambda\rightarrow \lambda/\kappa,\quad \eta\rightarrow\eta/\kappa.
\label{k}
\end{equation}
Thus, equations (10)-(13), and (15) hold for the orbitally excited case,
with the replacements $\lambda\rightarrow \lambda/\kappa$ and
$\eta\rightarrow\eta/\kappa$.
The radial wave function is $f(r)$ is given by
\begin{equation}
f(r)=r^{\kappa b-1}exp\left[-a\left(r +\frac{1}{m}\int S(r)dr\right)\right].
\label{f}
\end{equation}
Note that the constant $b$ used in this paper differs from the $b$ defined by
Eq.\ (22) of Ref.\ 1.  The $b$ in this paper is the $b$ in Ref.\ 1 divided by
$\kappa$.

In summary, we have presented a class of potentials for which the
Dirac equation has relatively simple exact analytical solutions for the ground
state of each angular momentum state with $l=j-\frac{1}{2}$.

\end{document}